\documentclass[aps,prl,reprint,superscriptaddress]{revtex4-2}

\usepackage{graphicx}
\usepackage{amsmath}
\usepackage{amssymb}
\usepackage{dcolumn}
\usepackage{bm}
\usepackage{multirow}
\usepackage{enumitem}
\usepackage{booktabs}
\usepackage{xcolor}
\usepackage{xspace}
\usepackage{bbold}
\usepackage{physics}
\usepackage[colorlinks=true, allcolors=blue]{hyperref}

\newcommand{\1}[0]{\widetilde{1}}

\begin{document}

\title{Freezing-Induced Interaction Engineering in Fixed-Frequency Transmon Networks}

\author{Songbo Xie}
\affiliation{Department of Electrical and Computer Engineering, North Carolina State University, Raleigh, North Carolina 27606, USA}

\author{Jiheng Duan}
\affiliation{Department of Physics and Astronomy, University of Rochester, Rochester, New York 14627, USA}
\affiliation{Center for Coherence and Quantum Science, University of Rochester, Rochester, New York 14627, USA}

\author{Sabre Kais}
\email{skais@ncsu.edu}
\affiliation{Department of Electrical and Computer Engineering, North Carolina State University, Raleigh, North Carolina 27606, USA}

\begin{abstract}
Freezing is usually used to suppress unwanted dynamics, but it can also be used to engineer interactions. We introduce freezing-induced interaction engineering, a drive-only control scheme in which a continuously driven auxiliary subsystem remains frozen while its drive-dependent state reshapes the effective Hamiltonian of neighboring degrees of freedom. We apply this idea to fixed-frequency transmon systems using a driven modulator coupled to a pivot transmon with fixed native exchange interactions to neighboring transmons. Changing only the modulator drive frequency controls the dressed resonance condition of the pivot, selectively activating a native interaction or detuning it from all neighbors. Full lab-frame simulations of multilevel transmons demonstrate this switching mechanism and show that it remains effective under finite-time frequency ramps. We further demonstrate the same control mechanism numerically in a four-transmon, two-neighbor setting, where changing only the modulator drive frequency selectively activates either neighboring native interaction or suppresses both. Repeating this local control unit along a fixed-frequency chain gives a scalable architecture for alternating layers of native iSWAP interactions.
\end{abstract}

\maketitle


{\it Introduction.---}Freezing is commonly used to suppress unwanted dynamics and protect quantum states~\cite{facchi2008quantum}. It can be realized through the quantum Zeno effect~\cite{misra1977zeno,itano1990quantum,schafer2014experimental,liu2025experimental}, dynamical decoupling~\cite{viola1998dynamical,viola1999universal,morton2006bang,uhrig2007keeping,damodarakurup2009experimental}, and spin locking~\cite{timoney2011quantum,bodey2019optical,zuk2024robust}. More generally, freezing constrains selected degrees of freedom and can change how the remaining system evolves. This principle has been explored in quantum control~\cite{facchi2004unification,pazsilva2012zeno}, interaction switching~\cite{zhou2009quantum}, projected-Hamiltonian engineering~\cite{burgarth2014exponential}, constrained quantum simulation~\cite{stannigel2014constrained}, and entangling operations~\cite{lewalle2023multi}. A frozen subsystem can therefore reshape the effective Hamiltonian governing the unfrozen system.

This Hamiltonian reshaping can be understood through interaction projection. If a subsystem $A$ is frozen in a state $|\psi\rangle$ while coupled to another subsystem $B$ through $V_{AB}$, the interaction is projected to $\langle\psi|_A V_{AB}|\psi\rangle_A$ acting on $B$, modifying its effective Hamiltonian~\cite{xie2025strong}. In our previous work, we used this mechanism to switch an always-on energy-transfer channel by turning a stroboscopic Zeno-kick sequence on and off~\cite{xie2026modulator}. Here, $A$ remains frozen under a continuous microwave drive. Varying the drive frequency changes the frozen state of $A$ and, through the projected interaction, modifies the effective Hamiltonian of $B$.

We apply this idea to interaction control in fixed-frequency transmon processors. Existing approaches can dynamically tune qubit frequencies~\cite{strauch2003quantum,dewes2012characterization,barends2014superconducting}, modify coupler-mediated interactions~\cite{blais2003tunable,bialczak2011fast,chen2014qubit}, or activate interactions with microwave drives applied to computational qubits~\cite{chow2011simple,chow2013microwave,magesan2020effective} or intermediate couplers~\cite{dipaolo2022extensible,shirai2023all,jiang2025microwave}. A closely related drive-based approach is off-resonant ac-Stark tuning, which directly dresses a computational qubit and shifts its transition frequency~\cite{blais2007quantum,zhao2022combating}. In contrast, we drive an auxiliary modulator outside the native interaction path between the computational transmons. The computational transmons remain undriven, their bare frequencies and native couplings stay fixed, and the auxiliary drive changes the resonance condition of the native interaction.

\begin{figure}[b]
\centering
\includegraphics[width=0.95\linewidth]{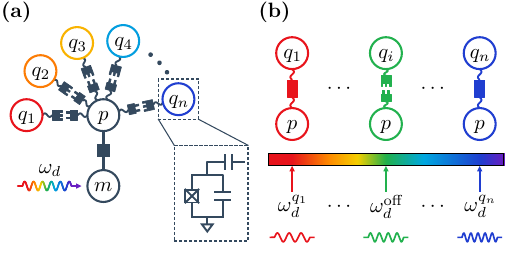}
\caption{
\textbf{Concept of freezing-induced interaction engineering.}
(a) A fixed-frequency pivot transmon $p$ is capacitively coupled to neighboring transmons $q_i$ $(i=1,2,\ldots,n)$ and to an auxiliary modulator transmon $m$. A coherent drive with carrier frequency $\omega_d$ is applied only to $m$, while the bare transmon frequencies and native couplings remain fixed.
(b) The drive-dependent frozen state of $m$ shifts the effective local Hamiltonian of $p$. At $\omega_d=\omega_d^{\mathrm{off}}$, the pivot is detuned from all neighboring transmons. At $\omega_d=\omega_d^{q_i}$, the pivot is selectively brought into resonance with $q_i$, activating the native $pq_i$ interaction while the remaining links stay off resonant. The colored arrows indicate representative drive frequencies for selecting different neighboring transmons.
}
\label{fig:system}
\end{figure}

For a concrete realization, we consider a local control unit consisting of a driven modulator $m$, a pivot transmon $p$, and neighboring transmons $q_i$. The modulator is coupled only to $p$, which has fixed native exchange interactions with the $q_i$. Freezing $m$ in a drive-dependent state modifies the effective local Hamiltonian of $p$. Changing only the drive frequency controls $p$'s dressed detuning from the neighboring transmons and selects which native interaction is resonant. The drive sets the resonance condition, while the native $pq_i$ coupling generates the iSWAP operation. Full lab-frame simulations of multilevel transmons demonstrate high-fidelity iSWAP operation, directly quantify the residual population dynamics in the off regime, and show that the gate remains robust under finite-time frequency ramps. We further demonstrate selective routing between multiple neighbors and show how the same local control unit can be repeated along a chain to form alternating layers of iSWAP gates.


{\it General framework.---}We model the architecture in Fig.~\ref{fig:system} as a pivot transmon $p$ coupled to neighboring transmons $\mathbf{q}\equiv\{q_i\}$, with a modulator transmon $m$ coupled only to $p$. Each transmon $j\in\{m,p,\mathbf{q}\}$ is described as a Duffing oscillator, $H_j=\omega_j a_j^\dagger a_j+(\alpha_j/2)a_j^{\dagger 2}a_j^2$, with $\alpha_j$ the anharmonicity. The full lab-frame Hamiltonian is (we use $\hbar=1$ throughout the work)
\begin{equation}\label{Htot}
    \begin{gathered}
        H_{mp\mathbf{q}}=\sum_{j\in\{m,p,\mathbf{q}\}}H_j+\Omega\cos(\omega_dt)X_m\\
        +J_{mp}(a_m+a_m^\dagger)(a_p+a_p^\dagger)+\sum_iJ_{pq_i}(a_p+a_p^\dagger)(a_{q_i}+a_{q_i}^\dagger).
    \end{gathered}
\end{equation}
The only time-dependent control is the transverse drive on $m$, with amplitude $\Omega$ and frequency $\omega_d$.


Moving to a frame rotating at $\omega_d$ and defining $\Delta_j\equiv\omega_j-\omega_d$, the single-transmon Hamiltonians become $\widetilde H_j=\Delta_j a_j^\dagger a_j+\alpha_j a_j^{\dagger 2}a_j^2/2$. Applying the rotating-wave approximation (RWA), the full Hamiltonian becomes (see Sec.~I of \cite{supplemental})
\begin{equation}\label{rotatingframe}
    \begin{gathered}
        H'_{mp\mathbf q}=H'_m(\omega_d)+\widetilde H_p+\sum_i\widetilde H_{q_i}\\
        +J_{mp}(a_m a_p^\dagger+a_m^\dagger a_p)
        +\sum_iJ_{pq_i}(a_p a_{q_i}^\dagger+a_p^\dagger a_{q_i}),
    \end{gathered}
\end{equation}
where $H'_m(\omega_d)\equiv\widetilde H_m+\tfrac{\Omega}{2}(a_m+a_m^\dagger)$ is the driven modulator Hamiltonian. We denote by $|F_m(\omega_d)\rangle$ the dressed state continuously connected to the bare state $|0_m\rangle$. Over the operating range considered here, it is the highest-energy eigenstate of $H'_m$ within the truncated modulator Hilbert space (see Sec.~I of \cite{supplemental}). When the relevant transitions from this state are sufficiently far detuned from the pivot transition, the $mp$ exchange is suppressed, and a modulator initialized in $|F_m(\omega_d)\rangle$ remains frozen up to perturbative corrections.

\begin{figure}[!t]
    \centering
    \includegraphics[width=0.92\linewidth]{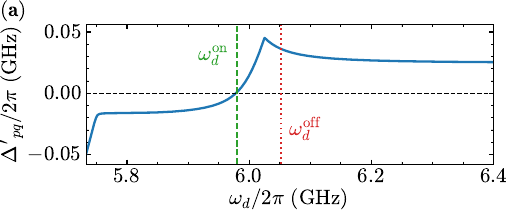}
    \includegraphics[width=0.46\linewidth]{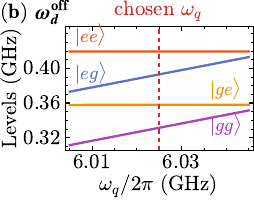}
    \includegraphics[width=0.46\linewidth]{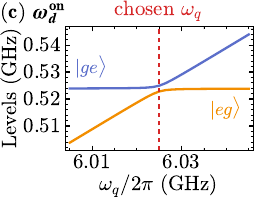}
    \caption{
    \textbf{Freezing-induced interaction engineering in the single-neighbor setting.} Unless varied in the corresponding panel, parameters are $\omega_m/2\pi=\omega_p/2\pi=6~\mathrm{GHz}$, $\omega_q/2\pi=6.025~\mathrm{GHz}$, $\alpha_m/2\pi=-350~\mathrm{MHz}$, $\alpha_p/2\pi=-500~\mathrm{MHz}$, $\alpha_q/2\pi=-350~\mathrm{MHz}$, $J_{mp}/2\pi=20~\mathrm{MHz}$, $J_{pq}/2\pi=1.5~\mathrm{MHz}$, and $\Omega/2\pi=600~\mathrm{MHz}$, with cutoff dimensions $d_m=5$, $d_p=4$, and $d_q=3$. 
    (a) Freezing-renormalized detuning $\Delta'_{pq}$ versus drive frequency $\omega_d$. The markers indicate the final dynamically calibrated operating frequencies. The interaction-on point $\omega_d^{\mathrm{on}}$ lies near the projected resonance $\Delta'_{pq}=0$, while $\omega_d^{\mathrm{off}}$ lies in the broad large-detuning region near the maximum of $\Delta'_{pq}$. For large positive $\omega_d$, $\Delta'_{pq}$ approaches the bare detuning $\omega_q-\omega_p$.
    (b) Floquet quasienergy spectrum in the interaction-off regime as $\omega_q$ is swept. The four relevant branches are shown with the modulator frozen in $|F_m\rangle^{\mathrm{off}}$. At the final dynamically calibrated frequency $\omega_d^{\mathrm{off}}$, the relevant branches remain separated, suppressing $pq$ exchange. 
    (c) Floquet quasienergy spectrum in the interaction-on regime. At the final dynamically calibrated frequency $\omega_d^{\mathrm{on}}$, an avoided crossing between $|F_m g_p e_q\rangle^{\mathrm{on}}$ and $|F_m e_p g_q\rangle^{\mathrm{on}}$ appears at the chosen $\omega_q$, signaling resonant $pq$ exchange with a splitting induced by the native $pq$ coupling. Vertical dashed lines in (b) and (c) mark the chosen $\omega_q$.
    }
    \label{fig:levels}
\end{figure}

In this freezing limit, projecting the $mp$ interaction onto $|F_m(\omega_d)\rangle$ reduces the full $mp\mathbf{q}$ Hamiltonian to
\begin{equation}\label{hpqprime}
    H'_{p\mathbf{q}}(\omega_d)
    =H'_p(\omega_d)+\sum_i\widetilde H_{q_i}
    +\sum_iJ_{pq_i}(a_p a_{q_i}^\dagger+a_p^\dagger a_{q_i}),
\end{equation}
where
\begin{equation}\label{hpprime}
    H'_p(\omega_d)=\widetilde H_p
    +J_{mp}\left(\langle a_m\rangle a_p^\dagger+\langle a_m^\dagger\rangle a_p\right),
\end{equation}
with $\langle a_m\rangle\equiv\langle F_m(\omega_d)|a_m|F_m(\omega_d)\rangle$. The projection produces an effective static transverse field on the pivot. We denote by $|g_p(\omega_d)\rangle$ and $|e_p(\omega_d)\rangle$ the eigenstates of $H'_p$ continuously connected to the bare states $|1\rangle$ and $|0\rangle$. Within the truncated Hilbert space, these are the second-highest and highest eigenstates, respectively. Their splitting is $\omega'_p(\omega_d)$, while the undriven neighboring transmons retain $\omega'_{q_i}(\omega_d)=|\Delta_{q_i}|$.

Each native $pq_i$ interaction is then controlled by the dressed detuning
\begin{equation}\label{detuning_general}
    \Delta'_{pq_i}(\omega_d)
    =\omega'_p(\omega_d)-\omega'_{q_i}(\omega_d).
\end{equation}
A global-off operating point satisfies
\begin{equation}\label{switching_off}
    \left|\Delta'_{pq_i}(\omega_d^{\mathrm{off}})\right|\gg J_{pq_i},
    \qquad \forall i,
\end{equation}
while selective activation of neighbor $q_k$ requires
\begin{equation}\label{switching_selective}
    \Delta'_{pq_k}(\omega_d^{(k)})=0,
    \quad
    \left|\Delta'_{pq_j}(\omega_d^{(k)})\right|\gg J_{pq_j},
    \quad \forall j\neq k.
\end{equation}
When these operating points are spectrally resolved, varying only the modulator drive frequency selects a native $pq_i$ interaction or detunes the pivot from all neighboring transmons. 

Because the microwave drive acts only on $m$, the pivot and neighboring transmons are not directly exposed to the strong drive. This indirect drive allows relatively large $\Omega$ while keeping leakage from the pivot and neighboring transmons small (see results below).


\begin{figure}
    \centering
    \includegraphics[width=0.9\linewidth]{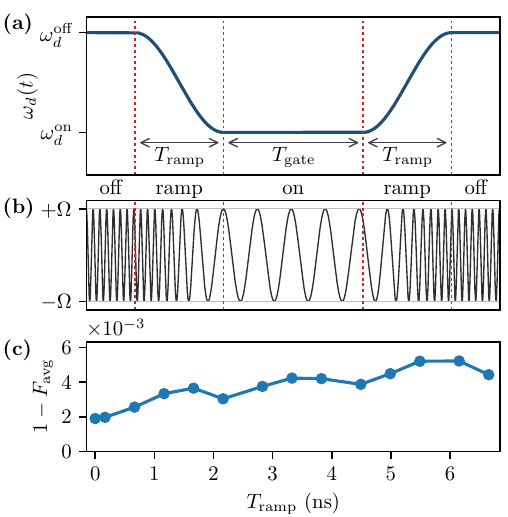}
    \caption{
    \textbf{Finite-time frequency switching and ramp-aware gate calibration.}
    (a) Instantaneous drive frequency $\omega_d(t)=\dot{\phi}(t)$ during an off--on--off switching sequence. The drive frequency is changed between $\omega_d^{\mathrm{off}}$ and $\omega_d^{\mathrm{on}}$ using half-cosine ramps of duration $T_{\mathrm{ramp}}$, with an interaction-on hold of duration $T_{\mathrm{gate}}$.
    (b) Corresponding phase-continuous drive $\Omega\cos[\phi(t)]$. The drive amplitude remains fixed throughout the sequence; the frequency contrast is exaggerated for visibility. Red dashed lines indicate the boundaries between the ramp and hold intervals.
    (c) Average gate infidelity after ramp-aware local corrections and recalibration of $T_{\mathrm{gate}}$, shown as a function of $T_{\mathrm{ramp}}$. The point at $T_{\mathrm{ramp}}=0$ corresponds to instantaneous switching. Markers denote the numerical results, and the solid line connects the data points.
    }
    \label{fig:ramp}
\end{figure}

For a single neighboring transmon $q$, the selective-control conditions reduce to an on/off switch of the native $pq$ interaction. Fig.~\ref{fig:levels}(a) shows the dressed detuning $\Delta'_{pq}(\omega_d)$ with the dynamically calibrated on and off points marked. Full lab-frame Floquet spectra evaluated at these same frequencies verify the two regimes (see Sec.~III of \cite{supplemental}). At $\omega_d^{\mathrm{off}}$, the relevant quasienergy branches remain separated, suppressing $pq$ exchange [see Fig.~\ref{fig:levels}(b)]. At $\omega_d^{\mathrm{on}}$, an avoided crossing between $|F_m g_p e_q\rangle^{\mathrm{on}}$ and $|F_m e_p g_q\rangle^{\mathrm{on}}$ appears, indicating resonant exchange with a splitting set by $J_{pq}$ [see Fig.~\ref{fig:levels}(c)]. Since $m$ is an auxiliary transmon, $|F_m(\omega_d)\rangle$ need not be confined to the bare $|0\rangle$ and $|1\rangle$ levels; higher-level population is part of the intended multilevel dressed state, and the relevant requirement is that $m$ remain frozen in $|F_m(\omega_d)\rangle$ during the operation.


{\it Numerical verification.---}We evaluate the protocol using the full lab-frame dynamics of the multilevel transmon system. In the interaction-on regime, gate performance is quantified by the average fidelity $F_{\mathrm{avg}}$ with respect to an ideal iSWAP, with population outside the active-pair computational subspace counted as error. In the interaction-off regime, we quantify the residual dynamics through the worst-case population loss. For each dressed computational-basis state $|\mathbf{z}\rangle$ of the pivot and its neighbors, let $P_\mathbf{z}(nT_d)$ be the probability of remaining in the initial configuration after $n$ drive periods, where $T_d=2\pi/\omega_d$. We define the stroboscopic worst-case population loss as
\begin{equation}
\mathcal L^{\mathrm{off}}
=
\max_\mathbf{z}\max_{0\leq n\leq N_{\mathrm{off}}}
\left[1-P_\mathbf{z}(nT_d)\right],
\label{eq:off_population_loss}
\end{equation}
where $N_{\mathrm{off}}$ is the nearest integer for which $N_{\mathrm{off}}T_d\simeq T_{\mathrm{ref}}$.

For selective routing, we quantify the disturbance of an inactive neighbor $q_j$ by its end-of-gate population loss $\mathcal{C}_{q_j}$. We also monitor leakage outside the dressed computational subspace of each transmon. Since the neighboring-transmon leakage is negligible, we report only the pivot leakage $\ell_p$. In the two-neighbor setting, the reported gate fidelity takes the worse of the two results obtained with the inactive neighbor initialized in $|g\rangle$ or $|e\rangle$. Detailed definitions and evaluation procedures are given in Sec.~IV of \cite{supplemental}, with Hilbert-space truncation and convergence discussed in Sec.~II. Throughout these numerical tests, we use representative device parameters without joint optimization and calibrate the operating frequencies and gate-level controls for each setting. Device-level optimization is left open and could further improve the reported performance.

We first consider a single neighbor. Using the representative device parameters in Fig.~\ref{fig:levels}, held fixed throughout the calculation, we calibrate the drive frequencies to $\omega_d^{\mathrm{on}}/2\pi=5.97966$ GHz and $\omega_d^{\mathrm{off}}/2\pi=6.04275$ GHz (see Sec.~III of \cite{supplemental}). At the interaction-on point, the full lab-frame dynamics implements an iSWAP in $T_{\mathrm{gate}}=234.96$ ns with $\mathcal{I}^{\mathrm{on}}=1.988\times10^{-3}$, corresponding to $F_{\mathrm{avg}}=99.801\%$, and pivot leakage $\ell_p^{\mathrm{on}}=0.074\%$. At the interaction-off point, using $T_{\mathrm{ref}}=T_{\mathrm{gate}}$, the worst-case population loss is $\mathcal{L}^{\mathrm{off}}=1.48\%$, while the maximum pivot leakage over the same time window is $\ell_p^{\mathrm{off}}=0.069\%$.

We implement the switch using phase-continuous half-cosine frequency ramps from $\omega_d^{\mathrm{off}}$ to $\omega_d^{\mathrm{on}}$, followed by an interaction-on hold and a reverse ramp [see Figs.~\ref{fig:ramp}(a) and (b)]. The drive amplitude remains fixed throughout the sequence, so the modulator is continuously driven. The finite ramps are incorporated into a calibrated sequence of local basis-conversion and correction operations whose nontrivial factors act on the modulator and pivot, with the neighboring transmon left unchanged. These operations convert between the dressed bases of the interaction-off and interaction-on operating points and compensate coherent local rotations generated during the ramps. The interaction-on hold time is also recalibrated for each ramp duration. The complete gate sequence and calibration procedure are given in Sec.~V of \cite{supplemental}. Full lab-frame simulations give $F_{\mathrm{avg}}$ between $99.479\%$ and $99.804\%$ for one-way ramp durations $T_{\mathrm{ramp}}=0.166$--$6.654~\mathrm{ns}$ [see Fig.~\ref{fig:ramp}(c)]. For the longest ramp, the calibrated infidelity is $4.421\times10^{-3}$, corresponding to $F_{\mathrm{avg}}=99.558\%$.

We next consider selective routing between two neighbors in the four-transmon architecture $m$--$p$--$\{q_1,q_2\}$. We take $\omega_1/2\pi=5.975$ GHz, $\omega_m/2\pi=\omega_p/2\pi=6.000$ GHz, $\omega_2/2\pi=6.025$ GHz, $J_{p1}/2\pi=J_{p2}/2\pi=1.5$ MHz, $J_{mp}/2\pi=32$ MHz, and $\Omega/2\pi=600$ MHz, with cutoff dimensions $(d_m,d_p,d_{q_1},d_{q_2})=(5,4,3,3)$. With these parameters fixed, calibration gives three operating points: $\omega_d^{q_1}/2\pi=6.04696$ GHz for activating $p$--$q_1$, $\omega_d^{q_2}/2\pi=5.88766$ GHz for activating $p$--$q_2$, and $\omega_d^{\mathrm{off}}/2\pi=6.00138$ GHz for suppressing both interactions (see Sec.~VI of \cite{supplemental}).

At $\omega_d^{q_1}$, the $p$--$q_1$ link implements an iSWAP in $T_{q_1}=200.60$ ns with $\mathcal{I}_{q_1}^{\mathrm{on}}=5.203\times10^{-3}$, corresponding to $F_{\mathrm{avg},q_1}=99.480\%$. The pivot leakage is $\ell_p^{q_1}=0.075\%$, while the inactive-neighbor crosstalk is $\mathcal{C}_{q_2}=0.429\%$. At $\omega_d^{q_2}$, the $p$--$q_2$ link implements an iSWAP in $T_{q_2}=187.51$ ns with $\mathcal{I}_{q_2}^{\mathrm{on}}=4.152\times10^{-3}$ and $F_{\mathrm{avg},q_2}=99.585\%$; the corresponding pivot leakage and inactive-neighbor crosstalk are $\ell_p^{q_2}=0.055\%$ and $\mathcal{C}_{q_1}=0.154\%$. At the common interaction-off point, using $T_{\mathrm{ref}}=\max(T_{q_1},T_{q_2})=200.60$ ns, the worst-case population loss is $\mathcal{L}^{\mathrm{off}}=1.67\%$, and the maximum pivot leakage is $\ell_p^{\mathrm{off}}=0.201\%$. These results show selective routing with a single driven modulator: changing only $\omega_d$ activates either neighboring native interaction or suppresses both. This two-neighbor unit also serves as the local building block for the scalable construction below, with the pivot leakage remaining no larger than $0.201\%$ for $\Omega/2\pi=600$ MHz.


\begin{figure}
    \centering
    \includegraphics[width=0.9\linewidth]{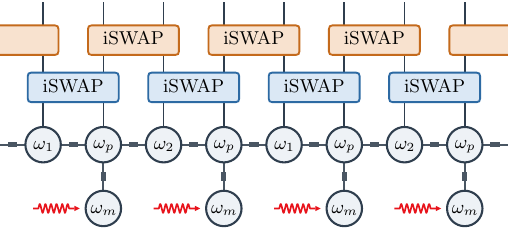}
    \caption{
    \textbf{Scalable architecture from repeated local control units.}
    A one-dimensional fixed-frequency chain $q_1$--$p_1$--$q_2$--$p_2$--$\cdots$ is formed with the non-pivot transmons alternating between $\omega_1$ and $\omega_2$, while each pivot transmon has frequency $\omega_p$ and is coupled to its own modulator with $\omega_m=\omega_p$. Each modulator is driven at one of three operating frequencies to select the left neighboring link, the right neighboring link, or the common interaction-off configuration. Because the local units are identical, the same three drive frequencies are reused throughout the chain. The two colored gate layers show the resulting brickwork pattern of parallel iSWAP operations on $(q_1,p_1),(q_2,p_2),\ldots$ and $(p_1,q_2),(p_2,q_3),\ldots$.
    }
    \label{fig:circuit}
\end{figure}

{\it Scalable architecture.---}The two-neighbor control unit above can be repeated to form larger fixed-frequency networks. As a simple example, consider a one-dimensional nearest-neighbor chain $q_1$--$p_1$--$q_2$--$p_2$--$\cdots$, with a modulator $m_j$ coupled to each pivot $p_j$ (see Fig.~\ref{fig:circuit}). All pivots have bare frequency $\omega_p$, while the non-pivot transmons $q_j$ alternate between $\omega_1$ and $\omega_2$. The modulators are chosen resonant with the pivots, $\omega_m=\omega_p$. Each pivot, its two neighboring transmons, and its attached modulator form the same four-transmon unit studied above. For each unit, the modulator is driven at one of three operating frequencies: two selectively activate the two neighboring links, while a third suppresses both. Since the local units have the same set of device parameters, these three drive frequencies can be reused throughout the chain. A chain of $N$ computational transmons requires $\lfloor N/2\rfloor$ modulators. In contrast, using one tunable coupler per nearest-neighbor link requires $N-1$ couplers.

Different modulators can be driven at different operating frequencies simultaneously, and no common rotating frame for all modulators is required. To leading order in the residual off-resonant couplings, embedding a local unit in the chain does not alter the switching conditions in Eqs.~\eqref{switching_off} and \eqref{switching_selective}. Unselected links remain far detuned, so their influence enters only as higher-order corrections to the local four-transmon dynamics (see Sec.~VII of \cite{supplemental}). For parallel operation, the activated links are chosen to be disjoint, so that no transmon participates in two gates simultaneously.

The chain can then be operated in alternating iSWAP layers. In one layer, each pivot selects its left neighbor, activating $(q_1,p_1),(q_2,p_2),\ldots$ in parallel. In the next layer, the pivots select their right neighbors, activating $(p_1,q_2),(p_2,q_3),\ldots$. Because the non-pivot transmons $q_j$ alternate between $\omega_1$ and $\omega_2$, the assignment of the two interaction-on drive frequencies to the left and right links alternates from one pivot to the next, while inactive units can be kept at the common interaction-off frequency. The resulting drive pattern produces the brickwork sequence shown in Fig.~\ref{fig:circuit} using repeated fixed-frequency control units.


{\it Summary.---}We have introduced a drive-only scheme for engineering native interactions in fixed-frequency transmon systems through dynamical freezing. A continuously driven auxiliary modulator remains frozen while its drive-dependent state reshapes the effective Hamiltonian of a coupled pivot transmon. Changing only the modulator drive frequency tunes the pivot into or out of resonance with neighboring transmons, while their bare frequencies and native couplings remain fixed. Full lab-frame simulations of multilevel transmons demonstrate iSWAP operation with $F_{\mathrm{avg}}=99.801\%$ in the single-neighbor setting and a worst-case interaction-off population loss of $1.48\%$. The switching remains effective under phase-continuous finite-time frequency ramps. With two neighbors, the same modulator selectively addresses either native interaction with fidelities of $99.480\%$ and $99.585\%$, with inactive-neighbor crosstalk no larger than $0.429\%$. These values are obtained for representative device parameters without joint optimization and serve as proof-of-principle operating points; further device-level optimization is left open. Repeating the local control unit along a fixed-frequency chain gives a scalable architecture for alternating layers of native iSWAP interactions.

{\it Acknowledgment.---}We thank Dr.~Zihao Wang for valuable suggestions and Dr.~Feiyang Ye for helpful discussions. We acknowledge support from the US Department of Energy, Office of Basic Energy Sciences, through the Quantum Photonic Integrated Design Center (QuPIDC) EFRC award DE-SC0025620.

\end{document}